\documentclass[twocolumn,prb,showpacs]{revtex4}
\usepackage{graphicx}
\usepackage{dcolumn}
\usepackage{bm}
\usepackage{color}

\begin{document}
\title{Dynamics in two-leg spin ladder with a four-spin cyclic interaction}
\author{S.~Nishimoto}
\affiliation{Leibniz-Institut f\"ur Festk\"orper- und Werkstoffforschung Dresden, D-01171 Dresden, Germany}
\author{M.~Arikawa}
\affiliation{Institute of Physics, University of Tsukuba 1-1-1 Tennodai, Tsukuba Ibaraki 305-8571, Japan}
\date{\today}
\begin{abstract}
We study two-leg Heisenberg ladder with four-spin cyclic interaction using 
the (dynamical) density-matrix renormalization group method. We demonstrate 
the dependence of the low-lying excitations in the spin wave, staggered dimer 
order, and scalar-chirality order structure factors on the four-spin cyclic 
interaction. We find that the cyclic interaction enhances spin-spin correlations 
with wave vector around momentum $(q_x,q_y)=(\frac{\pi}{2},0)$. Also, 
the presence of long-range order in the staggered dimer and scalar-chirality 
phases is confirmed by a $\delta$-function peak contribution of the structure 
factors at energy $\omega=0$.
\end{abstract}
\maketitle 

For many years, it had been generally believed that the magnetic 
properties of undoped high-$T_{\rm c}$ materials can be well described 
by the two-dimensional (2D) Heisenberg model with nearest-neighbor 
exchange interaction $J$. However, four-spin cyclic interaction $K$ 
has been increasingly recognized as a non-negligible correction to 
the Heisenberg model. The cyclic interaction comes from the forth-order 
processes in the strong-coupling limit of the single-band Hubbard model 
at half filling.~\cite{Takahashi77} The importance of this interaction was 
initially proposed in the 2D solid $^3$He, which has the hard-core correlations 
between spin-$\frac{1}{2}$ fermions.~\cite{Roger83}

In fact, a substantial value $K=0.24J$ was proposed for 2D copper 
oxide La$_2$CuO$_4$ by an accurate fit of the magnon dispersion.~\cite{Katanin02} 
A close value was also suggested by an analysis of the Raman-scattering 
data.~\cite{Honda93} Such magnitude of the four-spin cyclic interaction 
must have a considerable influence at least quantitatively on the low-energy 
spin physics. Similar situations have been reported for several two-leg 
spin-ladder systems: the exchange interactions were estimated as 
$J_\parallel=J_\perp=110$ meV and $K=16.5$ meV for 
La$_6$Ca$_8$Cu$_{24}$O$_{41}$ (Ref.~\onlinecite{Matsuda00}); 
$J_\parallel=186$ meV, $J_\perp=124$ meV, and $K=31$ meV for 
La$_4$Sr$_{10}$Cu$_{24}$O$_{41}$ (Ref.~\onlinecite{Notbohm07}); 
$J_\parallel=165$ meV, $J_\perp=150$ meV, and $K=15$ meV for 
SrCu$_2$O$_3$ (Ref.~\onlinecite{Mizuno99}), where $J_\parallel$ 
and $J_\perp$ are exchange interactions in the leg and rung 
directions, respectively. 

Motivated by those observations, the ground-state properties of 
two-leg spin-$\frac{1}{2}$ Heisenberg ladder with the four-spin cyclic 
interaction have been intensively studied.~\cite{Lauchli03,Hikihara03,Schmidt03,Hijii03,Notbohm07} 
Also, the effect of magnetic field on the ground state has been 
investigated.~\cite{Nakasu01,Hikihara08} Furthermore, the spectral 
features of the spin structure factor have been examined by 
the exact diagonalization, perturbation theory, and density-matrix 
renormalization group (DMRG) method.~\cite{Brehmer99,Haga02,Nunner02,Inaba05} 
The spin dynamics for small cyclic interactions is thus well understood, 
but the dynamical properties for other correlations and/or relatively 
large cyclic interactions are still open. In this paper, we study 
the dynamical structure factors of staggered dimer order, 
scalar-chirality order, and spin waves for a wide range of the 
four-spin cyclic interaction to give a deeper insight into our knowledge 
of the low-lying excitations, using the dynamical DMRG (DDMRG) 
method.~\cite{Jeckelmann02}

\begin{figure}[t]
    \includegraphics[width= 6.0cm,clip]{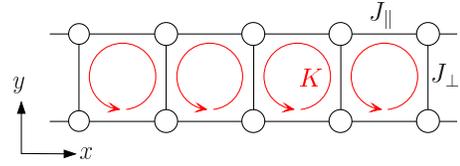}
  \caption{
Lattice structure of the two-leg Heisenberg ladder. $J_\parallel$ ($J_\perp$) 
is the exchange interaction in the leg (rung) direction and $K$ is 
the four-spin cyclic interaction. The $x$- ($y$-) axis is defined as the leg (rung) 
direction.
  }
    \label{fig1}
\end{figure}

The Hamiltonian of the two-leg spin-$\frac{1}{2}$ Heisenberg ladder with 
the four-spin cyclic interaction is given by
\begin{eqnarray}
\nonumber
H &=& J_\parallel \sum_{x,y} \vec{S}_{x,y} \cdot \vec{S}_{x+1,y} 
+ J_\perp\sum_{x} \vec{S}_{x,1} \cdot \vec{S}_{x,2} \\
&&+ K \sum_x (P_x+P_x^{-1}),
\label{hamiltonian}
\end{eqnarray}
with the cyclic permutation operator
\begin{eqnarray}
\nonumber
P_x+P_x^{-1}&=&\vec{S}_{x,1} \cdot \vec{S}_{x,2}+\vec{S}_{x+1,1} \cdot \vec{S}_{x+1,2}
+\vec{S}_{x,1} \cdot \vec{S}_{x+1,1}\\
\nonumber
&&+\vec{S}_{x,2} \cdot \vec{S}_{x+1,2}
+\vec{S}_{x,1} \cdot \vec{S}_{x+1,2}+\vec{S}_{x,2} \cdot \vec{S}_{x+1,1}\\
\nonumber
&&+4(\vec{S}_{x,1} \cdot \vec{S}_{x,2})(\vec{S}_{x+1,1} \cdot \vec{S}_{x+1,2})\\
\nonumber
&&+4(\vec{S}_{x,1} \cdot \vec{S}_{x+1,1})(\vec{S}_{x,2} \cdot \vec{S}_{x+1,2})\\
&&-4(\vec{S}_{x,1} \cdot \vec{S}_{x+1,2})(\vec{S}_{x,2} \cdot \vec{S}_{x+1,1})
\label{cyclicterm}
\end{eqnarray}
where $\vec{S}_{x,y}$ is a spin-$\frac{1}{2}$ operator at a site $(x,y)$ 
[see Fig.~\ref{fig1}]. For simplicity, we focus on the case of $J_\parallel=J_\perp=J$ 
and take $J=1$ as the unit of energy hereafter. The ground-state phase diagram 
was obtained in Ref.~\onlinecite{Lauchli03} as follows: the system has a rung singlet 
phase for $-3.33 \lesssim K \lesssim 0.23$, a staggered dimer long-range order (LRO) 
phase for $0.23 \lesssim K \lesssim 0.5$, a scalar-chirality LRO phase for 
$0.5 \lesssim K \lesssim 2.8$, a dominant vector chirality phase for $2.8 \lesssim K$, 
and a ferromagnetic phase for $K \lesssim -3.33$.

Let us define the dynamical structure factor as
\begin{eqnarray}
\nonumber
A(\vec{q},\omega) = \sum_\nu 
\langle \psi_0 \big|\hat{\cal O}_{-\vec{q}}\big| \psi_\nu \rangle
\langle \psi_\nu \big|\hat{\cal O}_{\vec{q}}\big| \psi_0 \rangle \\
\times \delta(\omega-E_\nu+E_0),
\label{dynm}
\end{eqnarray}
where $| \psi_\nu \rangle$ is the $\nu$-th eingenstate with the eigenenergy $E_\nu$ 
and $\hat{\cal O}_{\vec{q}}$ is the Fourier transformation of the quantity-dependent 
operator $\hat{\cal O}_{\vec{r}}$. The $\delta$-function is replaced by a Lorenzian 
with width $\eta$ in our numerical calculations. We now study the following three 
kinds of the dynamical structure factor corresponding to three phases at 
$-3.3 \lesssim K \lesssim 2.8$. The first is spin structure factor $S(\vec{q},\omega)$ 
with the operator
\begin{eqnarray}
\hat{\cal O}_{\vec{r}}=S^z_{x,y},
\label{spindef}
\end{eqnarray}
where $S^z_{x,y}$ is $z$-component of the total spin, the second is dimer order structure 
factor $D(\vec{q},\omega)$ with
\begin{eqnarray}
\hat{\cal O}_{\vec{r}}=\vec{S}_{x-1,y} \cdot \vec{S}_{x,y}-\vec{S}_{x,y} \cdot \vec{S}_{x+1,y},
\label{dimerdef}
\end{eqnarray}
and the third is scalar-chirality structure factor $C(\vec{q},\omega)$ with
\begin{eqnarray}
\hat{\cal O}_{\vec{r}}=\vec{S}_{x,1} \cdot (\vec{S}_{x+1,1} \times \vec{S}_{x+1,2}).
\label{schidef}
\end{eqnarray}
By integrating Eq.~(\ref{dynm}), we can easily obtain the static structure factor
\begin{eqnarray}
A(\vec{q}) =  
\langle \psi_0 \big|\hat{\cal O}_{-\vec{q}} \hat{\cal O}_{\vec{q}}\big| \psi_0 \rangle.
\label{stat}
\end{eqnarray}

We employ the DDMRG method~\cite{Jeckelmann02} which is an extension 
of the standard DMRG method.~\cite{White92} It has been developed for 
calculating dynamical correlation functions at zero temperature in quantum 
lattice models. This method has been successfully applied to the 
one-dimensional Heisenberg model.~\cite{Nishimoto07} We now calculate 
the dynamical structure factor (\ref{dynm}) with applying the periodic 
boundary conditions in the leg ($x$) direction. We fix the system length 
$L=32$ and $\eta=0.1$ if not otherwise stated. In the DDMRG calculation, 
a required CPU time increases rapidly with the number of the density-matrix 
eigenstates so that we would like to keep it as few as possible; meanwhile, 
the DDMRG approach is based on a variational principle so that we have 
to prepare a `good trial function' of the ground state with the 
density-matrix eigenstates as much as possible. Therefore, we keep 
$m=600$ to obtain true ground state in the first ten DDMRG sweeps 
and keep $m=300$ to calculate the spectral functions. In this way, the maximum 
truncation error, i.e., the discarded weight, is about $3 \times10^{-4}$, 
while the maximum error in the ground-state and low-lying excited states 
energies is about $10^{-2}$. 

\begin{figure}[t]
    \includegraphics[width= 5.5cm,clip]{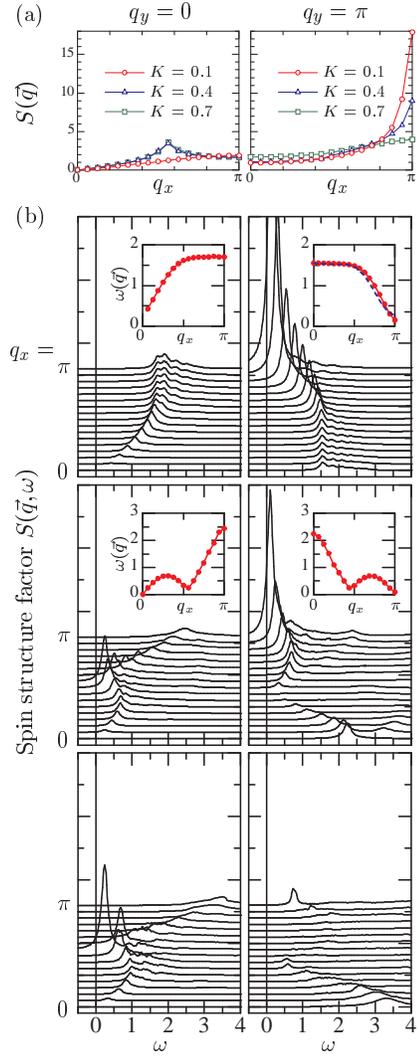}
  \caption{
(a) Static spin structure factor. (b) Dynamical spin structure factor for $K=0.1$ 
(Top), $K=0.4$ (Middle), and $K=0.7$ (Bottom). Left and right panels correspond 
to the results for $q_y=0$ and $q_y=\pi$, respectively. Insets: lower-edge of the 
two-spinon continuum ($q_y=0$) and one-spinon dispersion ($q_y=\pi$). The dashed 
line denotes the perturbative result 
$\omega(q_x,q_y=\pi)=1.186+0.558\cos(q_x)-0.271\cos(2q_x)+0.071\cos(3q_x)$.
  }
    \label{spec_S}
\end{figure}

To begin with, we consider the spin structure factor. In Fig.~\ref{spec_S}, 
we show the DMRG results of the static and dynamical spin structure 
factors for the rung singlet ($K=0.1$), staggered dimer LRO ($K=0.4$), 
and scalar-chirality LRO ($K=0.7$) phases. The dispersion relations 
$\omega(\vec{q})$ are also plotted in the insets. The spectra for 
$q_y=0$ and $\pi$ exhibit the two-triplon and one-triplon contributions, 
respectively. In the absence of the cyclic interaction, i.e., 
$K=0$,~\cite{Barnes93,Eder98} it is known that the spin dispersion has two 
minima at $q_x=0, \pi$ and a maximum at $q_x \sim 2\pi/3$ for $q_y=0$; whereas, 
two minima at $q_x=0, \pi$ and a maximum at $q_x \sim \pi/3$ for $q_y=\pi$. 
Those features have been confirmed to remain qualitatively unchanged at 
$K \lesssim 0.075$.~\cite{Nunner02} For $K=0.1$, however, the minima at 
$(q_x,q_y)=(\pi,0), (0,\pi)$ are no longer visible [see the insets of the 
top panels in Fig.~\ref{spec_S}(b)], i.e., the dispersions are nearly flat 
around $(q_x,q_y) \sim (\pi,0), (0,\pi)$. The one-triplon excitation ($q_y=\pi$) 
is in good agreement with the perturbative result.~\cite{Brehmer99} 
This is also consistent with other numerical study.~\cite{Haga02}

When the cyclic interaction is further increased to $K=0.4$, we can see 
a drastic change of the spectra for both $q_y=0$ and $\pi$: especially, 
an enhancement of peaks around $(q_x,q_y)=(\frac{\pi}{2},0)$ and 
a reduction of peaks around $(q_x,q_y)=(\pi,\pi)$ are derived. 
It is because the cyclic interaction leads to a repulsive interaction 
between neighboring rung triplets. In addition, a node emerges at 
$q_x=\frac{\pi}{2}$ for both $q_y$ values and a relation 
$\omega(q_x,q_y=0)=\omega(\pi-q_x,q_y=\pi)$ appears to be satisfied. 
They would indicate a two-fold degenerate ground state with a broken 
translational symmetry, which is consistent with the staggered dimer 
order state. For $K=0.7$, the peaks around $(q_x,q_y)=(\frac{\pi}{2},0)$ 
are still more enhanced and, whereas, the low-energy spectral features 
for $q_y=\pi$ seem to be much reduced. Actually, the one-triplon 
contribution is shunted off to the high-energy excitations since 
the static structure factor for $q_y=\pi$ is not much suppressed. 
In short, from the standpoint of spin-spin correlation the four-spin 
cyclic interaction may work for enhancing a spin-density wave with 
wave vector $(q_x,q_y)=(\frac{\pi}{2},0)$ and for reducing the 
antiferromagnetic correlation $S(\pi,\pi)$ [see Fig.~\ref{spec_S}(a)].

\begin{figure}[t]
    \includegraphics[width= 5.5cm,clip]{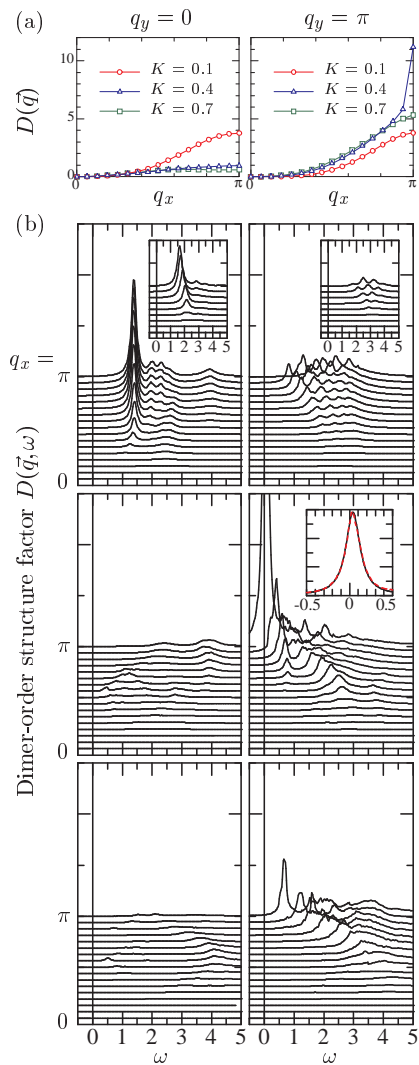}
  \caption{
(a) Static dimer-order structure factor. (b) Dynamical dimer-order structure 
factors for $K=0.1$ (Top), $K=0.4$ (Middle), and $K=0.7$ (Bottom). Left and 
right panels correspond to the results for $q_y=0$ and $q_y=\pi$, respectively.
Insets of the top panels: $D(\vec{q},\omega)$ for $K=0$ with $L=16$ and $\eta=0.2$.
Inset of the muddle panel: A Lorenzian fit of the peak at $(q_x,q_y)=(\pi,\pi)$ 
and $\omega \sim 0$ with $\eta=0.1$.
  }
    \label{spec_D}
\end{figure}

Next, we turn to the dimer-order structure factor. Figure~\ref{spec_D} 
shows the DMRG results of the static and dynamical dimer order structure 
factors for $K=0.1$, $0.4$, and $0.7$. For comparison, the results of 
$D(\vec{q},\omega)$ for $K=0$ are shown in the insets of the top panels 
of Fig.~\ref{spec_D}(b). In the rung-singlet phase, the ground state 
is approximately expressed as the product of local rung singlets with gap 
$\Delta \sim {\cal O}(J_\perp)$. The lowest excitation comes from 
the formation of a leg singlet with coupling energy $\sim \frac{J_\parallel}{2}$ 
as well as the collapse of two rung singlets. For $K=0$, therefore, 
undispersive sharp peaks appear around 
$\omega = {\cal O}(2\Delta-\frac{J_\parallel}{2}) \sim 1.5$ for $q_y=0$; 
whereas, the spectra for $q_y=\pi$ consist of broad continua at 
$\omega > {\cal O}(2\Delta-J_\parallel)$.

When small cyclic interaction ($K=0.1$) is introduced, we can see 
a strong influence on the continua around $(q_x,q_y)=(\pi,\pi)$, i.e., 
they are significantly shifted towards lower energies. It implies that 
the gap $\Delta$ is reduced rapidly as $K$ increases. For $K=0.4$, 
the continua are further drastically changed: a pronounced peak 
appears at $(q_x,q_y)=(\pi,\pi)$ and $\omega \sim 0$; also, most of the 
spectral weight concentrates around the peak. On the other hand, 
the spectral weights for $q_y=0$ are totally suppressed. The pronounced 
peak is well fitted by a Lorenzian with $\eta=0.1$, as shown in 
the inset of the middle panel of Fig.~\ref{spec_D}(b). In other words, 
the spectrum for $(q_x,q_y)=(\pi,\pi)$ may consist of a $\delta$-function peak 
at $\omega = 0$ and a gapfull continuum. They would be a signature 
of long-range staggered dimer order. For $K=0.7$, the spectral weights 
around $(q_x,q_y)=(\pi,\pi)$ are much reduced and a gap opens. 
It means that the staggered dimer order is no longer dominant in 
the ground state. Nevertheless, the spectral weights around 
$(q_x,q_y)=(\pi,\pi)$ are still significant, as seen in Fig.~\ref{spec_D}(a), 
so that the dimer-order correlation could just be changed from 
long-range order to short-range order. It is consistent with the 
fact that the staggered dimer order parameter is finite even in 
the scalar-chirality LRO phase.~\cite{Lauchli03}

\begin{figure}[t]
    \includegraphics[width= 8.0cm,clip]{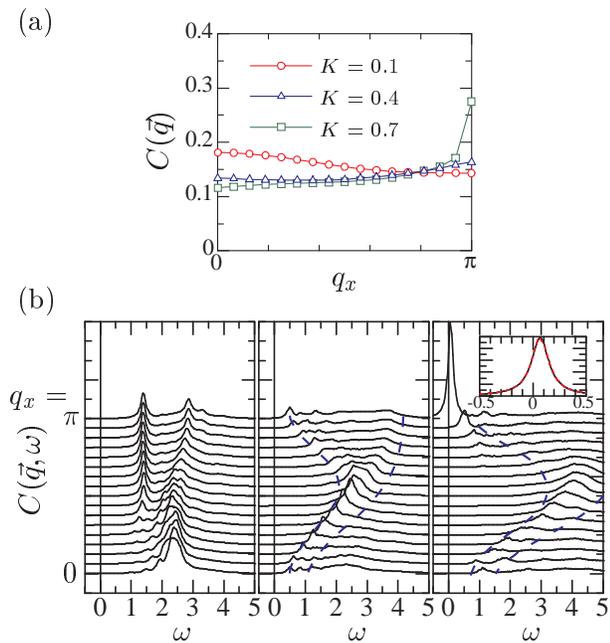}
  \caption{
(a) Static scalar-chirality structure factor. (b) Dynamical scalar-chirality 
structure factors for $K=0.1$ (left), $K=0.4$ (center), and $K=0.7$ (right). 
Inset of the bottom panel: A Lorenzian fit of the peak at $q_x=\pi$ and 
$\omega \sim 0$ with $\eta=0.1$. The dashed lines denote the lower and 
upper edges of the continuum.
  }
    \label{spec_C}
\end{figure}

Finally, we look at the scalar-chirality structure factor. The DMRG results 
of the static and dynamical scalar-chirality structure factors for $K=0.1$, 
$K=0.4$, $K=0.7$ are shown in Fig.~\ref{spec_C}. For $K=0.1$, the lowest 
excitations are described by (almost) undispersive peaks around 
$\omega \sim {\cal O}(2\Delta-\frac{J_\parallel}{2})$ in analogy with the 
dimer-order structure factor. For $K=0.4$, the spectra form 
a continuum bounded by the branches $\omega(q_x) \sim A\sin(q_x)$ and 
$\omega(q_x) \sim 2A\sin(q_x/2)$, except that a gap opens at $q_x=0$ and 
$\pi$. The existence of the gap implies that the scalar-chirality order still 
belongs to an excited state. If we assume a complete staggered dimer order, 
i.e., the ground state is the product of local dimer singlets, 
the scalar-chirality operator (\ref{schidef}) may be effectively reduced as 
$\vec{S}_{x,1} \cdot (\vec{S}_{x+1,1} \times \vec{S}_{x+1,2}) \big| \psi_0 \rangle \approx (\vec{S}_{x,1}/2) \big| \psi_0 \rangle$. Thus, the dispersions are similar to those 
of the spin structure factor in the one-dimensional spin-Peierls Heisenberg 
model.~\cite{Tsvelik92,Haas95} For $K=0.7$, we can see the closing of the gap 
and, moreover, the appearance of a dominant peak at $q_x=\pi$ and $\omega \sim 0$. 
This peak is well fitted by a Lorenzian with $\eta=0.1$, as shown 
in the inset of the right panel of Fig.~\ref{spec_C}(b). Hence, 
the spectrum for $q_x=\pi$ is composed of a $\delta$-function peak at $\omega = 0$ 
and a gapfull continuum, as is $D(\vec{q},\omega)$ in the staggered dimer 
LRO phase. It must indicate the presence of the scalar-chirality LRO.

In summary, we study the two-leg Heisenberg ladder with the cyclic four-spin 
interaction. The static and dynamical structure factors for the spin waves, 
staggered dimer order, and the scalar-chirality order parameters are 
calculated with the DDMRG method. We find that the spin-spin correlation 
with wave vector $(q_x,q_y)=(\frac{\pi}{2},0)$ is enhanced by the cyclic 
interaction. We also confirm the presence of long-range order in the 
staggered dimer and scalar-chirality phases by a $\delta$-function peak 
contribution of the structure factor at $\omega=0$.

We thank M.~Nakamura, S.~Suga, T.~Tohyama, T.~Hikihara, T.~Momoi,  I.~Maruyama, S.~Tanaya and Y.~Hatsugai for useful discussions.  
This work has been supported in part by the University of 
Tsukuba Research Initiative and Grants-in-Aid for Scientific 
Research, No.20654034 from JSPS and No.220029004 (Physics 
of New Quantum Phases in Super-clean Materials) and 
No.20046002 (Novel States of Matter Induced by Frustration) 
on Priority Areas from MEXT for M.A.


\begin{thebibliography}{99}

\bibitem{Takahashi77} M.~Takahashi, J. Phys. C: Solid State Phys. {\bf 10}, 1289 (1977).
\bibitem{Roger83} M.~Roger, J.H.~Hetherington, and J.M.~Delrieu, \rmp {\bf  55}, 1 (1983).
\bibitem{Katanin02} A.A.~Katanin and A.P.~Kampf, \prb {\bf 66}, 100403(R) (2002).
\bibitem{Honda93} Y.~Honda, Y.~Kuramoto, and T.~Watanabe, \prb {\bf 47}, 11329 (1993).

\bibitem{Matsuda00} M.~Matsuda, K.~Katsumata, R.S.~Eccleston, S.~Brehmer and H.-J.~Mikeska, 
\prb {\bf 62}, 8903 (2000).
\bibitem{Notbohm07} S.~Notbohm, P.~Ribeiro, B.~Lake, D.A.~Tennant, K.P.~Schmidt, G.S.~Uhrig, 
C.~Hess, R.~Klingeler, G.~Behr, B.~Buchner, M.~Reehuis, R.I.~Bewley, C.D.~Frost, P.~Manuel, 
and R.S.~Eccleston, \prl {\bf 98}, 027403 (2007).
\bibitem{Mizuno99} Y.~Mizuno, T.~Tohyama, and S.~Maekawa, 
J. Low. Temp. Phys. {\bf 117}, 389 (1999).

\bibitem{Lauchli03} A.~L\"auchli, G.~Schmid, and M.~Troyer, \prb {\bf 67}, 100409(R) (2003).
\bibitem{Hikihara03} T.~Hikihara, T.~Momoi, and X.~Hu, \prl {\bf 90}, 087204 (2003).
\bibitem{Schmidt03} K.P.~Schmidt, H.~Monien, and G.S.~Uhrig, \prb {\bf 67}, 184413 (2003).
\bibitem{Hijii03} K.~Hijii, S.~Qin, and K.~Nomura, \prb {\bf 68}, 134403 (2003).

\bibitem{Nakasu01} A.~Nakasu, K.~Totsuka, Y.~Hasegawa, K.~Okamoto, and T.~Sakai, 
J. Phys.: Condens. Matter {\bf 13}, 7421 (2001).
\bibitem{Hikihara08} T.~Hikihara and S.~Yamamoto, J. Phys. Soc. Jpn. {\bf 77}, 014709 (2008).

\bibitem{Brehmer99} S.~Brehmer, H.-J.~Mikeska, M.~M\"uller, N.~Nagaosa, and S.~Uchida, 
\prb {\bf 60}, 329 (1999).
\bibitem{Haga02} N.~Haga and S.~Suga, \prb {\bf 66}, 132415 (2002).
\bibitem{Nunner02} T.S.~Nunner, P.~Brune, T.~Kopp, M.~Windt, and M.~Gr\"uninger, 
\prb {\bf 66}, 180404(R) (2002).
\bibitem{Inaba05} K.~Inaba and S.~Suga, Prog. Theor. Phys. Suppl. {\bf 159}, 128 (2005).

\bibitem{Jeckelmann02} E.~Jeckelmann, \prb {\bf 66}, 045114 (2002).

\bibitem{White92} S.R.~White, \prl {\bf 69}, 2863 (1992); \prb {\bf 48}, 10345 (1993).
\bibitem{Nishimoto07} S.~Nishimoto and M.~Arikawa, Int. J. Mod. Phys. B  {\bf 21}, 
2262 (2007).

\bibitem{Barnes93} T.~Barnes, E.~Dagotto, J.~Riera, E.S.~Swanson, \prb {\bf 47}, 3196 (1993).
\bibitem{Eder98} R.~Eder, \prb {\bf 57}, 12832 (1998).

\bibitem{Tsvelik92} A.M.~Tsvelik, \prb {\bf 45}, 486 (1992).
\bibitem{Haas95} S.~Haas and E.~Dagotto, \prb {\bf 52}, R14396 (1995).
 
\end{thebibliography}
\end{document}